# Intrinsic Layer-Polarized Anomalous Hall Effect in Bilayer MnBi$_2$Te$_4$

Rui Peng, Ting Zhang, Zhonglin He, Qian Wu, Ying Dai$^*$, Baibiao Huang, Yandong Ma$^*$

School of Physics, State Key Laboratory of Crystal Materials, Shandong University, Shandanan Street 27, Jinan 250100, China

*Corresponding author: daiy60@sdu.edu.cn (Y.D.); yandong.ma@sdu.edu.cn (Y.M.)

Layer-polarized anomalous Hall effect (LP-AHE) is an attractive phenomenon in condensed-matter physics from the standpoints of both fundamental interest and device applications. The current LP-AHE research is based on the extrinsic paradigm of using external electric fields, in which the generation and control of LP-AHE are not straightforward. Here, we propose a novel mechanism that realizes intrinsic LP-AHE in bilayer lattices, through the mediation of sliding physics and Berry curvature. Moreover, this mechanism could render the LP-AHE in a controllable and reversable fashion. We analyze the symmetry requirements for a system to host such intrinsic LP-AHE. Its validity is further demonstrated in a real material of bilayer MnBi$_2$Te$_4$. By stacking with broken inversion symmetry, the layer-locked Berry curvature enables the intrinsic LP-AHE in bilayer MnBi$_2$Te$_4$, and the switchable control of its LP-AHE is achieved by sliding ferroelectricity. Our work opens a significant new direction for LP-AHE and two-dimensional (2D) materials research.

Hall effect is an important electronic transport phenomenon, which can lead to novel physics and efficient device applications [1-5]. There are two different microscopic mechanisms for Hall effect:

one is extrinsic process arising from scattering effects, and the other is intrinsic mechanism related to Berry curvature [1-5]. By breaking either the inversion symmetry or time-reversal symmetry of materials, the large Berry curvature can arise because of the entangled Bloch bands with spin-orbit coupling (SOC). Two notable examples determined by this intrinsic contribution in 2D materials are the valley Hall effect (VHE) [6-10] and quantum anomalous Hall effect (QAHE) [11-15]: while the former is linked to inversion symmetry broken related valley degree of freedom [6-10], the latter connects with time-reversal symmetry broken related spin degree of freedom [11-15]. In the past decades, with the rise of 2D materials, the field of Berry curvature correlated Hall effects has undergone rapid development and received extensive attention at both fundamental and applied levels [16-18].

Recently, as a new member of Hall effect family, the LP-AHE is proposed in $MnBi_2Te_4$ thin films [19]. Instead of encoding with spin or valley degree of freedom, the LP-AHE rises through coupling Berry curvature to layer degree of freedom. It is regarded as a rich playground for new physics and offers unprecedented opportunities for developing next-generation information technologies [19]. The current LP-AHE research is established in the extrinsic paradigm of employing external electric field. However, for device applications, static fashions are preferred than dynamic strategies. The realization of intrinsic LP-AHE is thus of significant interest, but remains elusive. Meanwhile, different from the cases of VHE and QAHE [6-15], the LP-AHE is currently only realized in $MnBi_2Te_4$ thin films [19]. Actually, up to now, the LP-AHE has not yet been explored in monolayer or bilayer lattices. These facts pose an outstanding challenge for the field of LP-AHE research.

Here, based on the paradigm of mediating sliding physics and Berry curvature, we present a novel mechanism for realizing LP-AHE in bilayer lattices. This mechanism could enable the LP-AHE to occur intrinsically. And the obtained LP-AHE can be controlled and reversed through its coupling with sliding physics. The symmetry requirements for a system being with such intrinsic LP-AHE are mapped out. Furthermore, using first-principles calculations, we also show that such mechanism can be demonstrated in a real material of bilayer $MnBi_2Te_4$. Because of the broken inversion symmetry, the Berry curvature in bilayer $MnBi_2Te_4$ exhibits a layer-locked nature, which yields the intrinsic LP-AHE. Moreover, the LP-AHE in bilayer $MnBi_2Te_4$ shows a ferroelectric controllable fashion via its interlayer sliding. Our study thus not only provides a compelling mechanism toward the highly desired intrinsic LP-AHE, but also extends LP-AHE to 2D lattice.

**Methods**

First-principles calculations are performed based on density functional theory (DFT) [20] as implemented in Vienna ab initio simulation package (VASP) [21]. Exchange-correlation interaction is described by the Perdew-Burke-Ernzerhof (PBE) parametrization of generalized gradient approximation (GGA) [22]. We consider effective Hubbard $U_{eff}$ = 3.0 eV for *d* electrons of Mn atom [23-24]. Structures are relaxed until the force on each atom is less than 0.01 eV/Å. The cutoff energy and electronic iteration convergence criterion are set to 450 eV and $10^{-5}$ eV, respectively. To sample the 2D Brillouin zone, a Monkhorst–Pack (MP) k-grid mess [25] of 9 × 9 × 1 is used. To avoid the interaction between adjacent layers, a vacuum space of 20 Å is added. DFT-D3 method is employed to treat the van der Waals interaction [26]. Berry phase approach is employed to evaluate the vertical electric polarization [27], and ferroelectric (FE) switching pathway is obtained by nudged elastic band (NEB) method [28]. Berry curvature and anomalous Hall conductivity are calculated using the maximally localized Wannier functions (MLWFs) as implemented in WANNIER90 package [29]. We also employ VASPBERRY to calculate Berry curvature of the valence band [30].

**Results and discussion**

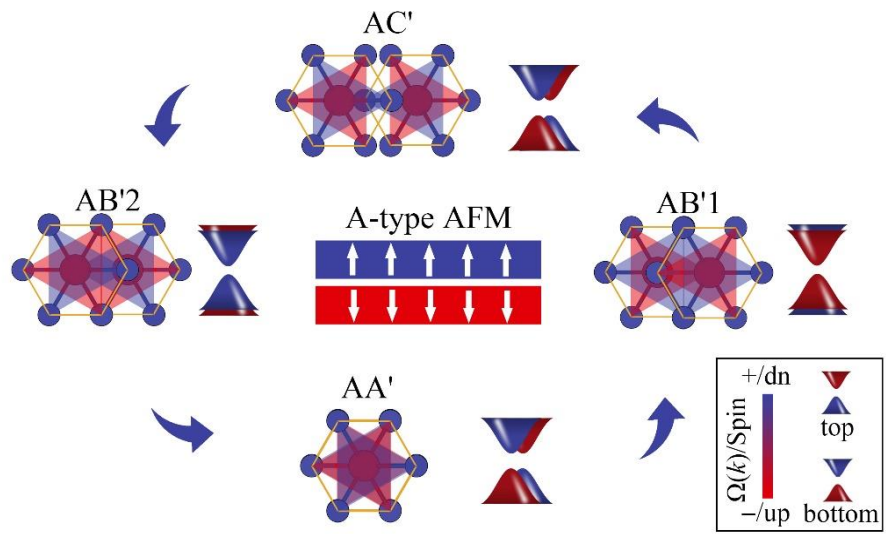

**Fig.1** Schematic diagrams of AA', AB'1, AB'2 and AC' configurations of a bilayer lattice and their band edges.

The mechanism is exemplified in a bilayer lattice. For the constituent single layers of such a bilayer lattice, two ingredients are essential. First, it should be a ferromagnetic semiconductor. Second, its band edges should have a large Berry curvature. With satisfying these two criteria, we stack them together to construct a bilayer lattice. For the bilayer lattice, two conditions are required: one is inversion/mirror symmetry breaking and the other is A-type antiferromagnetic (AFM) coupling [i.e., intralayer ferromagnetic (FM) coupling and interlayer AFM coupling].

To break the inversion symmetry, a rotation operation between the two layers is usually needed.

Here, we only focus on the single layers with a $C_{3z}$ rotation symmetry ($C_{3z}$-symmetry), which are ubiquitous in 2D lattices. In this case, one layer needs to rotate $(2m-1)\pi/3$ (m is a nonzero integer) with respect to the other. Without losing of generality, we set the rotation angle as $\pi$. To further break the mirror symmetry, an additional translation operation on one layer is required. In this regard, the top layer in the bilayer lattice can be stacked under the following symmetry operation on the bottom layer:

$$C_{2z}(0, 0, z)t(t_x, t_y, t_z).$$

Following this scenario, we obtain two high symmetric stacking patterns of the bilayer lattice, i.e., AB'1 $[C_{2z}(0, 0, z)t(a/3, -a/3, t_z)]$ and AB'2 $[C_{2z}(0, 0, z)t(-a/3, a/3, t_z)]$; see **Fig. 1**. Due to the symmetry breaking, an out-of-plane electric polarization pointing from bottom (top) to top (bottom) layers is induced in AB'1 (AB'2) configuration. Obviously, these two configurations can be regarded as two energetically equivalent FE states. Under the symmetry operations of $C_{2z}(0, 0, z)t(0, 0, t_z)$ and $C_{2z}(0, 0, z)t(a/2, -a/2, t_z)$, another two high symmetric stacking patterns can be achieved, which are referred to as AA' and AC', respectively; see **Fig 1**. Different from the cases of AB'1 and AB'2, the mirror and glide symmetries are preserved in AA' and AC', respectively. Therefore, the out-of-plane electric polarization is vanished in AA' and AC'. These two configurations can be considered as the nonpolar intermediate states, which corresponds to two ferroelectric transition pathways between AB'1 and AB'2, as shown in **Fig. 1**.

Different from the symmetry requirement, the latter condition is easy to be satisfied since A-type AFM coupling generally dominates the interlayer exchange interaction in bilayer systems [31-32]. Due to A-type AFM coupling, the band edges of bilayer lattice would have layer-locked Berry curvature. As shown in **Fig. 1**, the Berry curvatures of band edges contributed by the top and bottom layers are opposite. In AA' configurations, the bands are layer-degenerate under the protection of $M_z$-symmetry. While for AB'1 configuration, arising from the existence of an out-of-plane electric polarization, the degeneracy of spin-up bands contributed by the bottom layer and spin-down bands contributed by the top layer is lifted. Upon shifting the Fermi level between valence (conduction) band edges contributed by the two layers, the holes (electrons) from bottom (top) layer would feature an anomalous velocity in the presence of an in-plane electric field:

$$v \sim E \times \Omega(k)$$

As a result, the holes (electrons) would accumulate at one edge of bottom (top) layer. Under the ferroelectric transition, AB'1 configuration can be switched to AB'2 configuration, wherein the holes

(electrons) would accumulate at the other edge of top (bottom) layer. This scenario is also applicable for the case where there are band inversions between two layers in bilayer systems. This successfully establishes the intrinsic LP-AHE in the bilayer lattice.

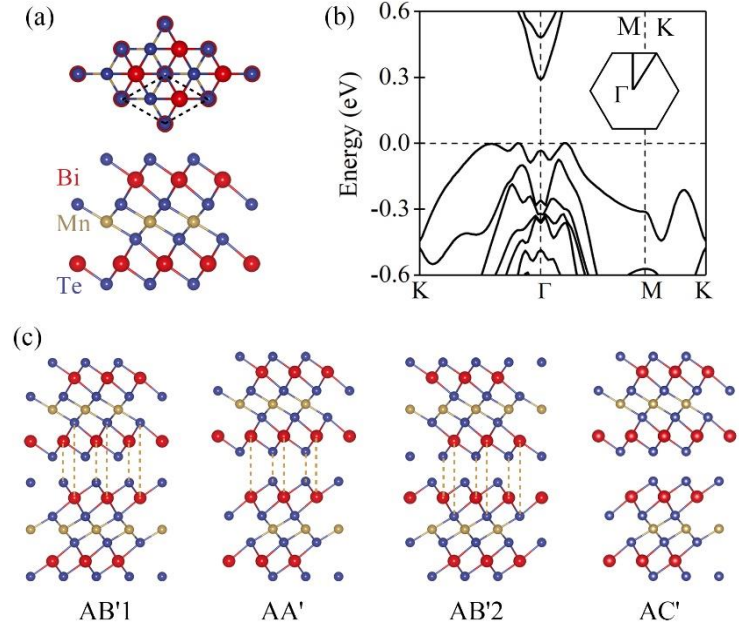

**Fig. 2** (a) Crystal structures of single-layer MnBi$_2$Te$_4$ from and side views. (b) Band structures of single-layer MnBi$_2$Te$_4$ with SOC. The Fermi level is set to 0 eV. (c) Crystal structures of bilayer MnBi$_2$Te$_4$ under AB'1, AA', AB'2 and AC' configurations.

After establishing the mechanism of intrinsic LP-AHE in bilayer lattice, we next discuss its realization in a real material of MnBi$_2$Te$_4$. **Fig. 2(a)** presents the crystal structures of single-layer MnBi$_2$Te$_4$. It exhibits a trigonal lattice with the space group of P3m1. Therefore, it has a C$_{3z}$ rotation symmetry. The lattice constant of single-layer MnBi$_2$Te$_4$ is optimized to be 4.31 Å. The valence electronic configuration of an isolated Mn atom is 3d$^5$4s$^2$. After donating two electrons to the six coordinated Te atoms, the left five electrons would half-fill the Mn-$d$ orbitals and thus result in a magnetic moment of 5 μ$_B$. Our first-principles calculations show the magnetic moment per unit cell is 5 μ$_B$, and magnetic moment on Mn is 4.51 μ$_B$. The exchange interaction among the magnetic moments favors FM coupling. **Fig. 2(b)** shows the band structure of single-layer MnBi$_2$Te$_4$ with SOC. We can see that single-layer MnBi$_2$Te$_4$ is a semiconductor with an indirect band gap of 0.29 eV. The valence band minimum (CBM) is located at the Γ point, while the valence band maximum (VBM) sits along Γ-M line. These results are consisted with the previous studies [23-24]. **Fig. S1** shows the Berry curvature of single-layer MnBi$_2$Te$_4$. Large Berry curvature can be observed at the Γ point.

Therefore, the two essential ingredients for constituent single layers analyzed above are satisfied.

Under the operation of $C_{2z}(0, 0, z)t(t_x, t_y, t_z)$, one single-layer MnBi$_2$Te$_4$ is stacked on the other, forming AB'1, AB'2, AA' and AC' configurations; see **Fig. 2(c)**. Among these four configurations, AB'1 and AB'2 have the lowest energy. To confirm their stabilities, the phonon spectra are calculated. As shown in **Fig. S2**, no imaginary frequency is found in the whole Brillouin zone for AB'1 and AB'2, indicating their dynamical stabilities. While for AA' and AC', there is a tiny imaginary frequency around the Γ point. To determine the magnetic ground states of these four systems, we consider different magnetic configurations. The energy differences between A-type AFM and FM coupling are -0.57, -0.57, -0.71 and -0.02 meV for AB'1, AB'2, AA' and AC', respectively. This indicates that A-type AFM coupling [**Fig. S3(a)**] is preferred for all these systems, satisfying the requirement for exchange interaction of the bilayer lattice. For AB'1 (AB'2) configuration, the magnetic moment on Mn atoms of the top layer in AB'1 (AB'2) is slightly smaller than that of the bottom layer, giving rise to a net magnetic moment of $7\times10^{-4}$ ($-7\times10^{-4}$) μ$_b$/unit cell. We also calculate the magnetocrystalline anisotropy energy (MAE) of AB'1 (AB'2). As shown in **Fig. S3(b)**, the out-of-plane magnetization is 0.004 meV more stable than in-plane magnetization. Therefore, AB'1 (AB'2) favors out-of-plane magnetization. Different from these two cases, the magnetic moment on Mn atoms of the top layer cancels out that of the bottom layer in AA' (AC') configuration, forming a net magnetic moment of 0 μ$_b$.

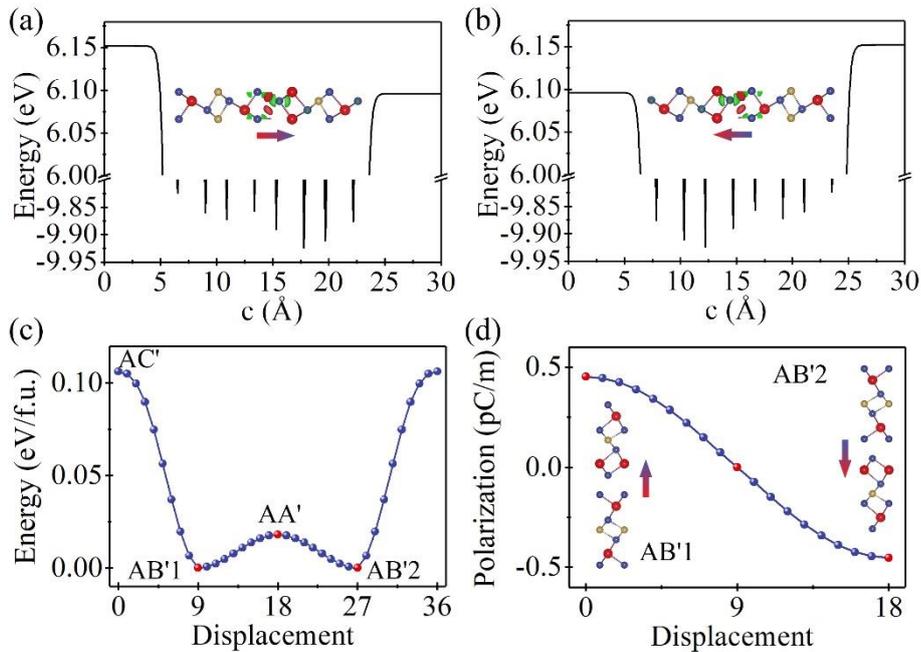

**Fig. 3** Plane averaged electrostatic potentials of (a) AB'1 and (b) AB'2 configurations for bilayer MnBi$_2$Te$_4$ along the *c* direction. Insets in (a) and (b) present the corresponding charge density of AB'1 and AB'2, respectively. Red and green isosurfaces represent electron accumulation and depletion,

respectively. (c) Energy profiles for FE switching of bilayer MnBi$_2$Te$_4$. (d) Variation of out-of-plane electric polarization of bilayer MnBi$_2$Te$_4$ under FE switching from AB'1 to AB'2.

At the interface of AB'1 configuration, the Te atom in the bottom layer is below the Bi atom in the top layer. This rises an out-of-plane electric polarization. To confirm the electric polarization, we calculate its charge density difference with respect to the two constituent single layers. As shown in **Fig. 3(a)**, there is a redistribution for charge density in the space between the two layers, namely, the electron accumulation and depletion areas are close to top and bottom layers, respectively, indicating the out-of-plane electric polarization pointing from bottom to top. This feature is also confirmed by the plane-averaged electrostatic potential of AB'1. As shown in **Fig. 3(a)**, there is a positive discontinuity ($\triangle V$ = 0.056 eV) between the vacuum levels of the top and bottom layers, again suggesting that AB'1 configuration exhibits an out-of-plane electric polarization pointing upward. Using the Berry phase approach, the electric polarization is calculated to be 0.45 pC/m for the AB'1 configuration, which is comparable to the recently reported FE vdW multilayers [33-35]. For AB'2 configuration, it is energetically equivalent to AB'1 and can be obtained from AB'1 under a mirror operation with respect to the horizontal plane. Therefore, the charge redistribution as well as $\triangle V$ are opposite to that of AB'1 configuration [**Fig. 3(b)**]. This results in an out-of-plane electric polarization of -0.45 pC/m in AB'2 configuration. To estimate the feasibility of FE transition between AB'1 and AB'2 configurations of bilayer MnBi$_2$Te$_4$, we calculate the energy barrier by using NEB method. As discussed above, AA' and AC' configurations can be considered as the nonpolar intermediate states, corresponding to two FE transition pathways between AB'1 and AB'2. As shown in **Fig. 3(c)**, the FE switching barrier along AC' is found to be 106 meV/f.u., while that along AA' is only 18 meV/f.u. Therefore, the FE transition is prone to occur along AA', and the corresponding electric polarizations as a function of step number is shown in **Fig. 3(d)**. It is worthy emphasizing that the switching barrier is comparable to that in the recently reported FE vdW multilayers [33-35], and that in In$_2$Se$_3$ (66 meV/f.u.) [36] and AgBiP$_2$Se$_6$ (6.2 meV/f.u.) [37]. Therefore, the FE switching in bilayer MnBi$_2$Te$_4$ harbors high feasibility.

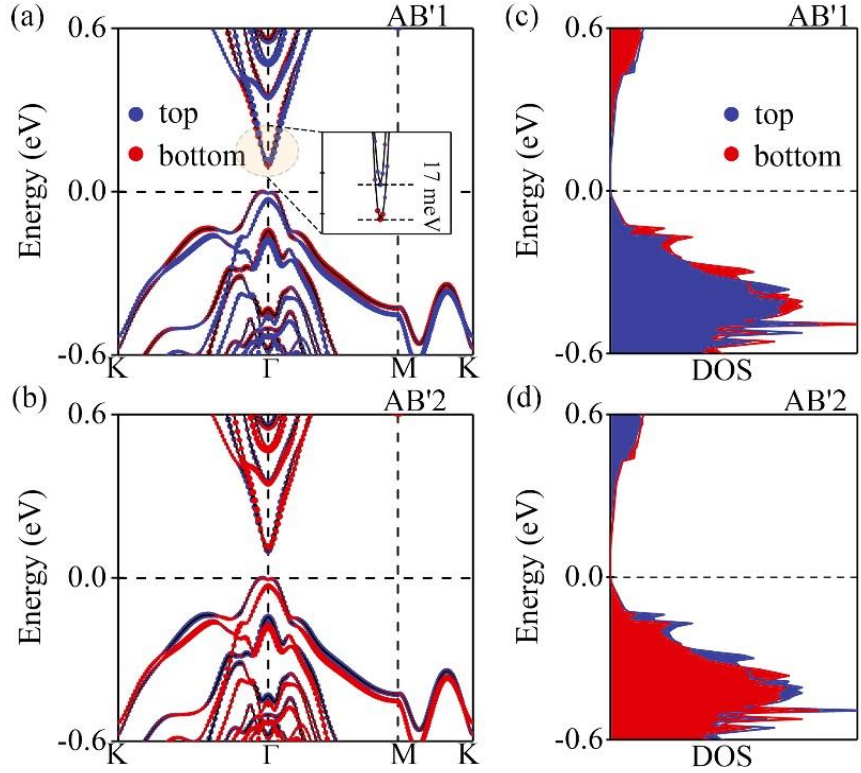

**Fig. 4** (a) Band structures and (c) density of states for AB'1 configuration of bilayer MnBi$_2$Te$_4$ with SOC. (b) Band structures and (d) density of states for AB'2 configuration of bilayer MnBi$_2$Te$_4$ with SOC. The Fermi level is set to 0 eV.

**Fig. 4** displays the band structures and density of states of AB'1 and AB'2 configurations. It can be seen that both systems exhibit an indirect band gap of 0.097 eV. The CBM is located at the Γ point, while the VBM sits along K-Γ line. The bands from top and bottom layers are separated for AB'1 and AB'2 configurations, although both systems exhibit AFM coupling. This feature can be attributed to the existence of an out-of-plane electric polarization. In addition, a band inversion is observed at Γ point. Correspondingly, for AB'1 (AB'2) configuration, the CBM is from the top (bottom) layer, while VBM is contributed by both layers. The separation between the bands from top and bottom layers can also be observed from the density of states shown in **Fig. 4(b)**. For comparison, we also calculate the band structure of AA'. As shown in **Fig. S3**, it has an indirect band gap of 0.20 eV. Its CBM is located at Γ point, while the VBM is along K-Γ line. The band structure is layer-degenerate along Γ-M line due to the protection of mirror symmetry, while there are layer splitting along K-Γ and M-K lines due to the effect of SOC and inversion symmetry breaking.

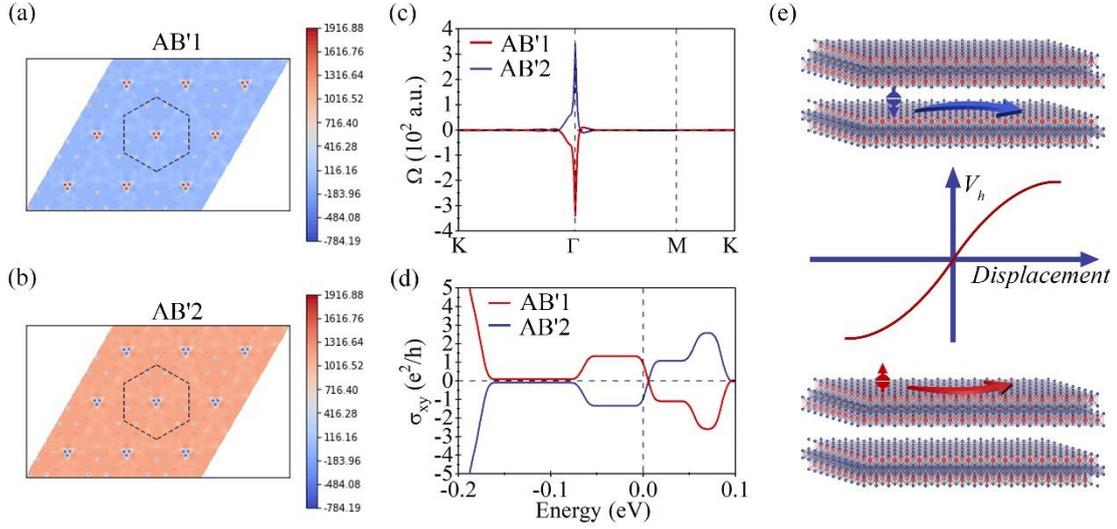

**Fig. 5** Berry curvature of the bottom conduction band as a contour map in the whole Brillion zone for (a) AB'1 and (b) AB'2 configurations of bilayer MnBi$_2$Te$_4$. (c) Berry curvatures along high symmetry points for AB'1 and AB'2 configurations of bilayer MnBi$_2$Te$_4$. (d) Anomalous Hall conductance for AB'1 and AB'2 configurations of bilayer MnBi$_2$Te$_4$. (e) Schematic diagrams of intrinsic LP-AHE in bilayer MnBi$_2$Te$_4$.

To explore the LP-AHE in bilayer MnBi$_2$Te$_4$, we calculate the Berry curvature of the bottom conduction band in AB'1 using the VASPBERRY code [30]. As shown in **Fig. 5(a)**, the Berry curvature of the bottom conduction band in AB'1 has large positive values around the Γ point. Accordingly, the electrons in the bottom conduction band around the Γ point for AB'1 would acquire an anomalous velocity $v \sim E \times \Omega(k)$, where E and $\Omega(k)$ are the external in-plane electric field and Berry curvature, respectively. When shifting Fermi level between the CBM and second bottom conduction band at the Γ point for AB'1, the electrons would be accumulated at the left edge of the bottom layer under an in-plane electric field [**Fig. 5(e)**], forming an anomalous Hall current. While for AB'2 configuration, the Berry curvature of the bottom conduction band exhibits large negative value around the Γ point, as shown in **Fig. 5(d)**. By shifting Fermi level between the CBM and second bottom conduction band at the Γ point for AB'2, the electrons would be accumulated at the right edge of the top layer under an in-plane electric field [**Fig. 5(e)**]. This also results in an anomalous Hall current. Therefore, the long-sought LP-AHE is realized in bilayer MnBi$_2$Te$_4$. Intriguingly, in sharp contrast to the previous work relying on external electric field [19], the LP-AHE occurs spontaneously based on the intrinsic electric polarization. And these two states can be easily switched to each other through ferroelectricity, as shown in **Fig. 5(e)**. Whereas in previous works on Hall effects, the reversal of anomalous velocity of carriers is through reversing the magnetization orientation, needing large energy-dissipating electric currents [6-15].

To confirm the LP-AHE in bilayer MnBi$_2$Te$_4$, we further investigate the anomalous Hall conductance (AHC), which is calculated using the following formula [38]:

$$\sigma = -\frac{e^2}{\hbar} \int \frac{d^2k}{(2\pi)^2} \Omega(k).$$

Here, the Berry curvature is calculated using the following formula [39]:

$$\Omega(k) = -\sum_n \sum_{n \neq n'} f_n \frac{2Im\langle\psi_{nk}|v_x|\psi_{n'k}\rangle\langle\psi_{n'k}|v_y|\psi_{nk}\rangle}{(E_n - E_{n'})^2}$$

Here, $f_n$ is the Fermi-Dirac distribution function, $\psi_{nk}$ is the Bloch wave function with eigenvalue $E_n$, and $v_{x/y}$ is the velocity operator along x/y direction. The calculated Berry curvature along the high symmetry points and AHC are plotted in **Fig. 5(c)** and **Fig. 5(d)**, respectively. Evidently, an AHC of ~1 (-1) e$^2$/h is obtained for the band around the CBM of AB'1 (AB'2). This firmly confirms the LP-AHE in bilayer MnBi$_2$Te$_4$.

**Conclusion**

In summary, a novel mechanism for intrinsic LP-AHE in bilayer lattices is proposed based on the paradigm of mediating sliding physics and Berry curvature. Through coupling with sliding physics, the proposed LP-AHE can be controlled and reversed via ferroelectricity. The symmetry requirements for hosting such intrinsic LP-AHE are mapped out. Using first-principles calculations, this mechanism is further demonstrated in a real material of bilayer MnBi$_2$Te$_4$. The Berry curvature in bilayer MnBi$_2$Te$_4$ displays a layer-locked nature, resulting in the LP-AHE intrinsically. And the LP-AHE in bilayer MnBi$_2$Te$_4$ is ferroelectric controllable via interlayer sliding. The proposed mechanism and the candidate bilayer MnBi$_2$Te$_4$ thus have the possibility of impacting LP-AHE research in 2D lattice.

**Acknowledgements**

This work is supported by the National Natural Science Foundation of China (No. 11804190, 12074217), Shandong Provincial Natural Science Foundation (Nos. ZR2019QA011 and ZR2019MEM013), Shandong Provincial Key Research and Development Program (Major Scientific and Technological Innovation Project) (No. 2019JZZY010302), Shandong Provincial Key Research and Development Program (No. 2019RKE27004), Shandong Provincial Science Foundation for Excellent Young Scholars (No. ZR2020YQ04), Qilu Young Scholar Program of Shandong University, and Taishan Scholar Program of Shandong Province.

**Reference**

[1] N. Nagaosa, J. Sinova, S. Onoda, A. H. MacDonald, and N. P. Ong, Anomalous Hall effect, Rev. Mod. Phys. 82, 1539 (2010).

[2] Z. Fang *et al.*, The anomalous Hall effect and magnetic monopoles in momentum space, Science 302, 92 (2003).

[3] F. D. M. Haldane, Berry curvature on the Fermi surface: Anomalous Hall effect as a topological Fermi-liquid property, Phys. Rev. Lett. 93, 206602 (2004).

[4] D. Xiao, M. C. Chang, and Q. Niu, Berry phase effects on electronic properties, Rev. Mod. Phys. 82, 1959 (2010).

[5] S. Nakatsuji, N. Kiyohara, and T. Higo, Large anomalous Hall effect in a non-collinear antiferromagnet at room temperature, Nature 527, 212 (2015).

[6] R. Peng, Y. Ma, X. Xu, Z. He, B. Huang, and Y. Dai, Intrinsic anomalous valley Hall effect in single-layer $Nb_3I_8$, Phys. Rev. B 102, 035412 (2020).

[7] R. Peng, Z. He, Q. Wu, Y. Dai, B. Huang, and Y. Ma, Spontaneous valley polarization in 2D organometallic lattice, Phys. Rev. B 104, 174411 (2021).

[8] Z. He, R. Peng, X. Feng, X. Xu, Y. Dai, B. Huang, and Y. Ma, Two-dimensional valleytronic semiconductor with spontaneous spin and valley polarization in single-layer $Cr_2Se_3$, Phys. Rev. B 104, 075105 (2021).

[9] W.-Y. Tong, S.-J. Gong, X. Wan, and C.-G. Duan, Concepts of ferrovalley material and anomalous valley Hall effect, Nat. Commun. 7, 13612 (2016).

[10] H. Hu, W. Tong, Y. Shen, X. Wan, and C. Duan, Concepts of the half-valley-metal and quantum anomalous valley Hall effect, npj Comput. Mater. 6, 129, (2020).

[11] F. D. M. Haldane, Model for a quantum Hall effect without Landau levels: Condensed-matter realization of the "Parity Anomaly", Phys. Rev. Lett. 61, 2015 (1988).

[12] K. He, Y. Wang, and Q. K. Xue, Quantum anomalous Hall effect, Nat. Sci. Rev. 1, 38 (2014).

[13] Z. Liu, G. Zhao, B. Liu, Z. Wang, J. Yang, and F. Liu, Intrinsic quantum anomalous Hall effect with in-plane magnetization: Searching rule and material prediction, Phys. Rev. Lett. 121, 246401 (2018).

[14] Z. He, R. Peng, Y. Dai, B. Huang, and Y. Ma, Single-layer $ScI_2$: A paradigm for valley-related multiple Hall effect, Appl. Phys. Lett. 119, 243102 (2021).

[15] Y. Jin, Z. Chen, B. Xia, Y. Zhao, R. Wang, and H. Xu, Large-gap quantum anomalous Hall phase in hexagonal organometallic frameworks, Phys. Rev. B 98, 245127 (2018).

[16] L. Smejkal, R. Gonzalez-Hernandez, T. Jungwirth, and J. Sinova, Crystal Hall effect in collinear antiferromagnets, Sci. Adv. 6, eaaz8809 (2020).

[17] N. J. Ghimire, A. S. Botana, J. S. Jiang, J. Zhang, Y.-S. Chen, and J. F. Mitchell, Large anomalous Hall effect in the chiral-lattice antiferromagnet $CoNb_3S_6$, Nat. Commun. 9, 3280 (2018).

[18] D. Shao, J. Ding, G. Gurung, S. Zhang, and E. Y. Tsymbal, Interfacial crystal Hall effect reversible by ferroelectric polarization, Phys. Rev. Appl. 15, 024057 (2021).

[19] A. Gao *et al.*, Layer Hall effect in a 2D topological axion antiferromagnet, Nature 595, 521 (2021).

[20] W. Kohn and L. J. Sham, Self-consistent equations including exchange and correlation effects, Phys. Rev. 140, A1133 (1965).

[21] G. Kresse and J. Furthmüller, Efficient iterative schemes for ab initio total-energy calculations using a plane-wave basis set, Phys. Rev. B 54, 11169 (1996).

[22] J. P. Perdew, K. Burke, and M. Ernzerhof, Generalized gradient approximation made simple, Phys. Rev. Lett. 77, 3865 (1996).

[23] Y. Li, Z. Jiang, J. Li, S. Xu, and W. Duan, Magnetic anisotropy of the two-dimensional ferromagnetic insulator $MnBi_2Te_4$, Phys. Rev. B 100, 134438 (2019).

[24] M. M. Otrokov, I. P. Rusinov, M. Blanco-Rey, M. Hoffmann, A. Yu. Vyazovskaya, S. V. Eremeev, A. Ernst, P. M. Echenique, A. Arnau, and E. V. Chulkov, Unique thickness-dependent properties of the van der Waals interlayer antiferromagnet $MnBi_2Te_4$ films, Phys. Rev. Lett. 122, 107202 (2019).

[25] H. J. Monkhorst and J. D. Pack, Special points for Brillouin-zone integrations, Phys. Rev. B 13, 5188 (1976).

[26] S. Grimme, J. Antony, S. Ehrlich, and H. Krieg, A consistent and accurate ab initio parametrization of density functional dispersion correction (DFT-D) for the 94 elements H-Pu, J. Chem. Phys. 132, 154104 (2010).

[27] R. D. King-Smith and D. Vanderbilt, Theory of polarization of crystalline solids, Phys. Rev. B 47, 1651 (1993).

[28] G. Mills, H. Jónsson, and G. K. Schenter, Reversible work transition state theory: Application to dissociative adsorption of hydrogen, Surf. Sci. 324, 305 (1995).

[29] A. A. Mostofi, J. R. Yates, G. Pizzi, Y.-S. Lee, I. Souza, D. Vanderbilt, and N. Marzari, An updated version of wannier90: A tool for obtaining maximally-localised Wannier functions, Comput. Phys. Commun. 185, 2309 (2014).

[30] H. Kim, C. Li, J. Feng, J. Cho, and Z. Zhang, Competing magnetic orderings and tunable topological states in two-dimensional hexagonal organometallic lattices, Phys. Rev. B 93, 041404(R) (2016).

[31] S. Gong, C. Gong, Y. Sun, W. Tong, C. Duan, J. Chua, and X. Zhang, Electrically induced 2D half-metallic antiferromagnets and spin field effect transistors, Proc. Natl. Acad. Sci. USA 115, 8511 (2018).

[32] H. Lv, Y. Niu, X. Wu, and J. Yang, Electric-field tunable magnetism in van der Waals bilayers with A‑type antiferromagnetic order: Unipolar versus bipolar magnetic semiconductor, Nano Lett. 21, 7050 (2021).

[33] T. Zhang, Y. Liang, X. Xu, B. Huang, Y. Dai, and Y. Ma, Ferroelastic-ferroelectric multiferroics in a bilayer lattice, Phys. Rev. B 103, 165420 (2021).

[34] Y. Liang, S. Shen, B. Huang, Y. Dai and Y. Ma, Intercorrelated ferroelectrics in 2D van der Waals materials, Mater. Horiz. 8, 1683 (2021).

[35] Q. Yang, M. Wu, and J. Li, Origin of two-dimensional vertical ferroelectricity in $WTe_2$ bilayer and multilayer, J. Phys. Chem. Lett. 9, 7160 (2018).

[36] W. Ding, J. Zhu, Z. Wang, Y. Gao, D. Xiao, Y. Gu, Z. Zhang, W. Zhu, Prediction of intrinsic two-dimensional ferroelectrics in $In_2Se_3$ and other $III_2$-$VI_3$ van der Waals materials, Nat. Commun. 8, 14956 (2017).

[37] B. Xu, H. Xiang, Y. Xia, K. Jiang, X. Wan, J. He, J. Yin, and Z. Liu, Monolayer $AgBiP_2Se_6$: an atomically thin ferroelectric semiconductor with out-plane polarization, Nanoscale, 9, 8427 (2017).

[38] Y. G. Yao, L. Kleinman, A. H. MacDonald, J. Sinova, T. Jungwirth, D.-S. Wang, E. Wang, and Q. Niu, First principles calculation of anomalous Hall conductivity in ferromagnetic bcc Fe, Phys. Rev. Lett. 92, 037204 (2004).

[39] D. J. Thouless, M. Kohmoto, M. P. Nightingale, and M. den Nijs, Quantized Hall conductance in a two-dimensional periodic potential, Phys. Rev. Lett. 49, 405 (1982).